\documentclass[conference]{IEEEtran}
\IEEEoverridecommandlockouts
\usepackage{cite}
\usepackage{amsmath,amssymb,amsfonts}
\usepackage{algorithmic}
\usepackage{textcomp}
\usepackage{xcolor}

\usepackage{graphicx}
\usepackage{multirow}
\usepackage{siunitx}
\usepackage[vlined,linesnumbered,ruled]{algorithm2e}
\makeatletter
\renewcommand{\@algocf@capt@plain}{above}
\makeatother
\usepackage{subcaption}
\usepackage[most]{tcolorbox}
\usepackage[export]{adjustbox}
\usepackage{array}
\usepackage{rotfloat}
\usepackage[T1]{fontenc}
\usepackage[latin9]{inputenc}
\usepackage{amssymb}
\usepackage{listings}
\usepackage{xcolor}

\definecolor{codegreen}{rgb}{0,0.6,0}
\definecolor{codegray}{rgb}{0.5,0.5,0.5}
\definecolor{codepurple}{rgb}{0.58,0,0.82}
\definecolor{backcolour}{rgb}{0.95,0.95,0.92}
 
\lstdefinestyle{code_style}{
    backgroundcolor=\color{backcolour},   
    commentstyle=\color{codegreen},
    keywordstyle=\color{magenta},
    numberstyle=\tiny\color{codegray},
    stringstyle=\color{codepurple},
    basicstyle=\ttfamily\tiny,
    breakatwhitespace=false,         
    breaklines=true,                 
    captionpos=b,                    
    keepspaces=true,                 
    numbers=left,                    
    numbersep=5pt,                  
    showspaces=false,                
    showstringspaces=false,
    showtabs=false,                  
    tabsize=2
}
 
\def\BibTeX{{\rm B\kern-.05em{\sc i\kern-.025em b}\kern-.08em
    T\kern-.1667em\lower.7ex\hbox{E}\kern-.125emX}}
\begin{document}

\title{Avocado: Open-Source Flexible Constrained Interaction Testing for Practical Application\\
{\footnotesize }
%\thanks{The authors acknowledge the support of the OP VVV funded project CZ.02.1.01/0.0/0.0/16\_019/0000765 "Research Center for Informatics".}
}

\author{\IEEEauthorblockN{Jan Richter}
\IEEEauthorblockA{Czech Technical University\\
Prague, Czech Republic \\
richtrja@gmail.com}
\and
\IEEEauthorblockN{Bestoun S. Ahmed}
\IEEEauthorblockA{Karlstad University\\
Karlstad, Sweden \\
bestoun@kau.se}
\and
\IEEEauthorblockN{Miroslav Bures}
\IEEEauthorblockA{Czech Technical University\\
Prague, Czech Republic \\
buresm3@fel.cvut.cz}
\and
\IEEEauthorblockN{Cleber R. Rosa Junior}
\IEEEauthorblockA{Red Hat, Inc. \\
Westford, USA \\
crosa@redhat.com}

}

\maketitle

\begin{abstract}
 
This paper presents the outcome of a research collaboration between academia and industry to implement and utilize the capabilities of constrained interaction testing for an open-source tool for industrial-scale application. The project helps promote flexibility in generating constrained interaction test suites, executing them, and setting up a test oracle to report them--all within the same tool called Avocado. Avocado employs a constraint solver with computational algorithms to generate constrained interaction test suites. The environment of the application under test can be set up to execute the generated test suite with minimum effort. A test oracle can be set up by the tool to report the status and the results of the executed test cases. Avocado represents a comprehensive and flexible solution for conducting combinatorial interaction testing (CIT) and constrained CIT on an industrial application. In this paper, we present the structure of the tool and our method of implementing the algorithms in detail.

\end{abstract}


\section{Introduction}

Combinatorial interaction testing (CIT) is a testing technique that relies on a mathematical object, covering array ($CA$), to represent the actual set of test cases based on the \textit{t-wise} coverage criteria (where $t$ represents the desired interaction strength of the input parameters or configurations of a software). $CA (N; t, k, v)$, also expressed as $CA (N; t, v^k)$, is a combinatorial structure that is constructed as an array of $N$ rows and $k$ columns on $v$ values such that every $N \times t$ sub-array comprises all ordered subsets from the $v$ values of size $t$ at least once \cite{KAMPEL2019107}. CA construction can be directly applied to \textit{t-wise} test-case reduction; therefore, considerable research has been conducted to develop effective strategies for test generation. However, in most real-world software applications, constraints are present among the input parameters. To this end, constrained CIT (CCIT) has emerged with the adoption of constrained CA (CCA) \cite{HARTMAN2004149,Jin2018}.

Over the past decade, in a move forward from the classical CIT, the research activities of CCIT have seen dramatic increase \cite{Yilmaz2014}. Similar to CIT, the CCIT research activities can be broadly classified into two main categories: generation of constrained interaction test suites and application of CCIT \cite{BestounSLS}. Over time, the research on test suite generation has produced excellent results, thereby rendering the generation type research mature. However, in recent years, an increasing interest has been observed in the application of CCIT owing to its excellent application in several industrial projects, e.g., \cite{Gargantini2013,Sherwood2015,Bozic2015,Li2016,Choi2016}. For practical application of CCIT, the tester must adapt it to the application environment. This adaptation uses the plugin of the test generation algorithm and involves the preparation of the test execution strategy. However, this concept being heavily dependent on the application type and its running environment, varies according to the application. To this end, each application requires custom setting, deployment, and configuration to apply CCIT. This is also true of CIT.

In this paper, we present a new approach to apply CCIT in practice through a unified framework, Avocado, that can assist the adaptation of CCIT to applications with minimum adaptation effort. Avocado is an automated testing framework that has been used in several successful industrial-scale projects. We implemented a CIT plugin that adds several capabilities to Avocado considering CCIT and achieves practical application of CCIT. Here, Avocado is able to generate CCIT test suites from a predefined set of application inputs and constraints among them, execute the generated test cases, and identify the test-cases' status based on a customized test oracle. In our earlier work \cite{Ahmed:2019}, we illustrated CIT with the Avocado testing framework that can be applied to a new application. In this paper, we present how Avocado handles the constraints and finally generates the combinatorial test suites. The paper also reports on how we applied the best practices from our works and other works published in literature to implement an open-source flexible tool with a practical application. Our main focus in this study is not to introduce new bounds of CAs or generated optimal test suites, compared to other tools. I. Hasan et. al \cite{Hasan2019GenerationAA} showed a detailed comparison of the test generation algorithm efficiency with other tools.

The rest of this paper is organized as follows. Section \ref{AvocadoFramework} presents an overview of the Avocado framework. Section \ref{CITVarianter} describes how the CIT varianter algorithm works within the Avocado framework. Section \ref{GenerationAlgo} provides the details of the test generation algorithm including constraint handling and optimization strategy. Finally, Section \ref{conclusion} concludes the paper.

\begin{figure*}
\centering
    \includegraphics[width= 4.7 in]{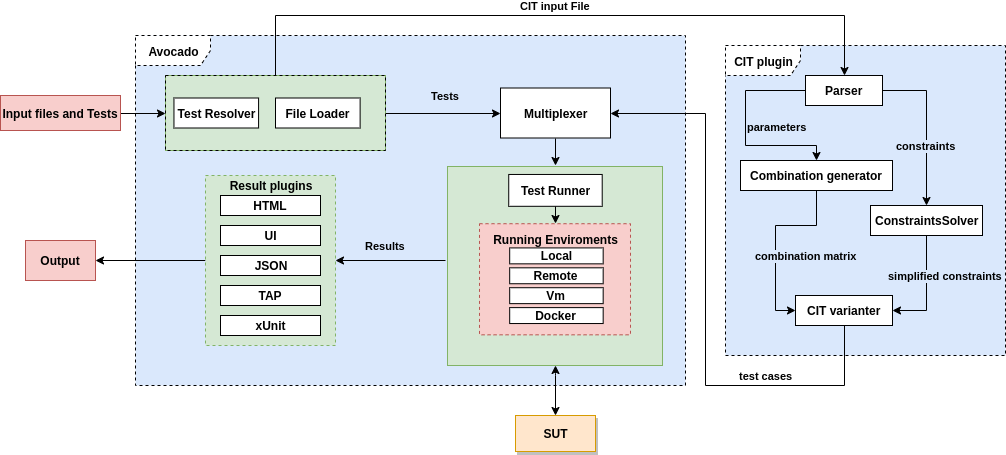}
    \caption{General structure of Avocado with the CIT plugin}
    \label{fig:Avocado_diagram}
\end{figure*}

\section{The Avocado framework}\label{AvocadoFramework}

Avocado\footnote{https://avocado-framework.github.io/} is an open-source testing framework that runs on Python, and it is maintained by Red Hat Inc. and the Avocado community contributors. It comprises a set of tools and libraries to facilitate the creation, execution, and evaluation of automated tests. It is the successor of the Autotest framework that was designed for the Linux kernel testing \cite{Admanski2010AutotestT}. It uses the best features of Autotest, while mitigating Autotests' weaknesses and drawbacks. The Avocado webpage contains more detailed information about those adopted features from Autotest. The native test cases are written in Python; therefore, they follow the unit-test pattern similar to a unit test in Python. However, Avocado has a utility for running any executable as a test.

As shown in Figure \ref{fig:Avocado_diagram}, Avocado comprises three main components: the test runner, libraries, and plugins. The test runner helps the users to execute test cases. Test cases can either be written in Python and use the Avocado API or in any language of the users' choice. In either cases, the Avocado runner's logging system records the activities during test execution to obtain important details about the software-under-test (SUT). The main APIs that Avocado makes available to test writers (known as "Test APIs"\footnote{https://avocado-framework.readthedocs.io/en/73.0/api/test/avocado.html} are derived and compatible with the Python unit test TestCase\footnote{https://docs.python.org/3/library/unittest.html\#unittest.TestCase} class, but adds many methods for functional and performance testing. Hence, the test runner is designed to help users run and log their test cases and brings a lot of features to make their work easier.

Avocado libraries facilitate users to write test cases in a concise, yet expressive and powerful manner. Libraries are the most important component of Avocado; they provide several tools to speed up and simplify test development. Plugins provide extensions to Avocado by facilitating addition of functionality and features to it.

Avocado has several plugins that can be divided into four groups: remote runner, result, variant, and testing plugins. Remote runner plugins enable users to run test cases remotely over SSH, using libvirt, or using Docker. Result plugins work with test results, and they can be used for presenting and formatting the output in different formats such as JSON, xUnit, HTML, and TAP, or saving results to a database or a dedicated server. Variant plugins create variants of the test data for running different variants of the tests in the Avocado test runner. Different varianters create different test data. The CIT varianter is a variant plugin that was recently developed by us; it is an excellent outcome of industry-academia collaboration.

Testing plugins can connect different testing frameworks to the Avocado framework. When a user requires certain features that are not present in the Avocado plugins, the solution is to create a customized plugin. In such a situation, the Avocado team creates a plugins system that simplifies the development process for new plugins. The user can then follow a few simple steps to create a plugin. These steps can be found in the Avocado documentation\footnote{https://avocado-framework.readthedocs.io}. The aforementioned utilities make Avocado a very powerful tool for automated testing and easily extensible to suit the varied needs of users.

\begin{figure}
%\scriptsize
    \begin{verbatim}
    PARAMETERS
    color[black,  gold, red]
    shape[square, triangle, circle]
    state[liquid, solid, gas]
    material[leather, plastic, aluminum]
    coating[anodic, cathodic]
    
    CONSTRAINTS
    color != black || shape != square
    color != black || shape != triangle
    color != black || shape != circle
    color != gold || coating != cathodic
    material != aluminum || color != gold
    \end{verbatim}
    \caption{CIT input file}
    \label{fig:cit_input_file}
\end{figure}

\section{CIT varianter plugin inside Avocado}\label{CITVarianter}

The CIT varianter plugin brings advantages of CIT and CCIT to the Avocado framework, wherein it can apply those testing methods in a fully automated testing process. The name varianter is Avocado-specific name and inspired by the variation of the test cases. For the test generation, the user must model the SUT parameters and constraints that will be included in the test cases. Avocado creates an input file in the CIT file format. Figure \ref{fig:cit_input_file} provides the structure of the input file. The first part describes the parameters of the SUT and its values. The second describes constraints among the parameters' values. The constraints must be in conjunctive normal form (CNF) and use the following three operands: \verb|!=, OR, AND|. Character || represents operand \verb|OR| and the new-line character represents operand \verb|AND|.

As illustrated in Figure \ref{fig:Avocado_diagram}, the CIT varianter generates from the input file, a set of test cases that comply with the input constraints and cover all $t-way$ combinations of the parameter values for a selected level of $t$. The $t-way$ combination of parameter values indicates that the varianter generates a set of all combinations of size T from the parameters for each value in the set. The CIT varianter computes all combinations of its parameter values. Further, this set of test cases is sent to the Avocado multiplexer that creates scripts for testing the SUT.

The multiplexer can obtain several inputs from different sources for different usage, e.g., statistical data, configuration files, and testing scripts. The multiplexer recognizes the test cases originating from the CIT varianter as \textit{variants}. When Avocado receives the data as variants, the test runner executes all test cases for each test case in the variants, records the output, and displays the verification message on the screen. Upon completion of the testing, the runner saves the recorded results in the format chosen by the user. The format can be XML, JSON, TAP, or HTML. The runner displays the output on the screen.

A significant feature of the Avocado CIT plugin is its customizable test oracle setting. Here, the user can configure the test oracle to handle the test cases automatically. Each test case is tagged by the runner as PASS or FAIL with additional information explaining the status. The status depends on the expected output of the test, and it can be customized by the user who can define the condition determining the PASS/FAIL status of the test. Here, the user can define the PASS/FAIL based on expected output. This can also be done by specifying conditions for a set of test cases rather than individual assignment of the output per each test case. Although this adds a great feature to Avocado, research and development is needed to support automation to this feature. The advantage of using the CIT plugin with Avocado is that the test runner can use the CIT and CCIT test cases for running tests in different environments and on machines that can be local, remote, or virtual. This was the description of how Avocado runs the test cases based on the CIT plugin. The following section illustrates in detail how the test generation algorithm generates test cases and deals with constraints.

%%%%%%%%%%%%%%% Bestoun 11 December

\section{Generating variants within Avocado}\label{GenerationAlgo}

The CIT plugin was designed, implemented, and maintained for flexible industrial usage. The aim was to use the capabilities of CIT and CCIT flexibly and practically. To this end, the CIT varianter uses a complex randomized algorithm of Monte Carlo type that finds a solution satisfying all the constraints in a reasonable time. Moreover, this solution need not be the optimal solution. Algorithm \ref{fig:full_pseudocode} provides a brief description of the working of this algorithm. The algorithm can be divided into four parts creating combinations, constraints processing, computing initial solution, and improvement of solution. The following subsections give the details of each part.

\begin{algorithm}

 \KwIn{parameters, constraints, tValue}
 \KwOut{Testcases}
 
     combinationMatrix = computeCombinations(data, tVvalue)\\
     solver = processConstraints(constraints, tValue)\\
     solver.cleanMatrix(combinationMatrix)\\
     solution = computeInitialSolution(combinationMatrix)\\
     \While{time != 0}{
     solution = computeBetterSolution(solution, combinationMatrix)\\
     }
     Return solution\\

    \caption{CIT varianter}
    \label{fig:full_pseudocode}
\end{algorithm}

\begin{comment}
\begin{figure}
\begin{lstlisting}
function compute_variants(data, constraints, t_value){

    combination_matrix = compute_combinations(data, t_value)
    solver = process_constraints(constraints, t_value)
    solver.clean_matrix(combination_matrix)
    solution = compute_initial_solution(combination_matrix)
    while(time != 0){
        solution = compute_better_solution(solution, combination_matrix)
    }
    return solution
}
\end{lstlisting}
    \caption{CIT varianter pseudocode}
    \label{fig:full_pseudocode}
\end{figure}
\end{comment}

%%%%%%%%%%%%%%% Bestoun 11 December

The following sections describes those parts of the generation algorithms in detail.

\subsection{Generating combinations}

The aim of the varianter within the CIT plugin is to generate test cases that cover all $t-way$ combinations of parameter values. To achieve this, the plugin needs to know the combinations and which test cases cover them. Consequently, the varianter creates a combination matrix with all $t-way$ combinations of the parameters' values (also called t-tuples). When the varianter obtains the input file, it processes the values as numbers in a sorted matrix. Here, indexing numbers are used for the input file entries from 0 to n. For example, the parameters in Figure \ref{fig:cit_input_file} are computed as \verb|color=0, shape=1 state=2|, and the values of the color parameter would be \verb|black=0, gold=1, red=2|. 

%%%%%%%%%%%%%%% Bestoun 11 December

The combination matrix uses the representation of the number of input parameters with all t-tuples. Each row of the combination matrix represents one combination of $t$ parameters and each column represents one combination of parameter values from a row. The value of a combination matrix cell represents how many test cases from the solution cover the exact t-tuples of parameter values. This cell can take values from -1 to the number of test cases, where -1 indicates that the combination does not exist. Figure \ref{fig:combination_matrix_initialization} illustrates the combination matrix of parameters from Figure \ref{fig:cit_input_file} (when t = 2) after initialization.

\begin{figure}
\scriptsize
    \bordermatrix{\text{ }&(0,0)&(0,1)&(0,2)&(1,0)&(1,1)&(1,2)&(2,0)&(2,1)&(2,2)\cr
                    (0,1)& 0 & 0 & 0 & 0 & 0 & 0 & 0 & 0 & 0\cr
                    (0,2)& 0 & 0 & 0 & 0 & 0 & 0 & 0 & 0 & 0\cr
                    (0,3)& 0 & 0 & 0 & 0 & 0 & 0 & 0 & 0 & 0\cr
                    (0,4)& 0 & 0 &-1 & 0 & 0 &-1 & 0 & 0 &-1\cr
                    (1,2)& 0 & 0 & 0 & 0 & 0 & 0 & 0 & 0 & 0\cr
                    (1,3)& 0 & 0 & 0 & 0 & 0 & 0 & 0 & 0 & 0\cr
                    (1,4)& 0 & 0 &-1 & 0 & 0 &-1 & 0 & 0 &-1\cr
                    (2,3)& 0 & 0 & 0 & 0 & 0 & 0 & 0 & 0 & 0\cr
                    (2,4)& 0 & 0 &-1 & 0 & 0 &-1 & 0 & 0 &-1\cr
                    (3,4)& 0 & 0 &-1 & 0 & 0 &-1 & 0 & 0 &-1\cr}
    \caption{Combination matrix initialization in Avocado CIT varianter}
    \label{fig:combination_matrix_initialization}
\end{figure}

\subsection{Constraints handling}

The CIT varianter also handles constraints entered in the input file. A constraint can be violated, thereby leading to a set of invalid and non-executable test cases. Figure \ref{fig:cit_input_file} shows an example of constraints. It can be seen that these constraints forbid the test cases with black color and square shape, black color and triangle shape, etc. However, constraints in real-world scenarios such as those on the industrial scale can be considerably complicated, and the varianter must process them efficiently. Moreover, the format to represent these constraints in the input file must be CNF \cite{WARNERS199863} for the ease of constraint computation and use. However, despite the CNF format of the constraints, the varianter must preprocess them to simplify them or find some implicit constraints that are not explicitly mentioned in the input file but are defined by other constraints. For example, in Figure \ref{fig:cit_input_file}, the first three constraints define that the color cannot be black because every shape cannot have black color. This means that we can reduce these three constraints to one. 

%%%%%%%%%%%%%%% Bestoun 12 December

It is not practical to let the user specify all the constraints in the SUT. However, it is possible to generate all the implicit constraints based on a small number of explicit ones. For preprocessing the constraints, the varianter implemented the constraint handling mechanism for handling forbidden tuples that are presented in \cite{7107441}. To transform the input constraints into forbidden tuples, we consider the CNF because each disjunction part of CNF represents a forbidden tuple. For example, we can transform the constraints in Figure \ref{fig:cit_input_file} into a set of five forbidden tuples as shown in Figure \ref{fig:tuples_input_file}. The varianter derives a set of forbidden tuples and uses it for constraints validation. A test case is valid if and only if it does not contain any forbidden tuple.

These properties work precisely with the data representation in the combination matrix. In other words, every combination that contains a forbidden tuple is forbidden. Based on this mechanism, the varianter finds the forbidden cells inside the combination matrix and tags them as -1, thereby indicating the "does not exist combination." This process is called matrix cleaning. However, before the varianter cleans the combination matrix, it must compute the forbidden tuples from the input constraints. For this computation, the varianter uses two methods: \textbf{derive}, for finding new implicit tuples from the existing set and \textbf{simplify}, for the reduction of unnecessary tuples from the set. 
%%%%%%%%%%%%%%% Bestoun 12 December

The \textbf{derive} method uses one rule to derive the implicit forbidden tuples from existing ones. If we have parameter \textit{P} with values \textit{n} then there are \textit{n} forbidden tuples, where each tuple contains different values of \textit{P}. The varianter constructs a new forbidden tuple by combining all values in these \textit{n} tuples except the values of parameter \textit{P}. Using this rule, the \textbf{derive} method finds all forbidden tuples for each parameter. Furthermore, if the rule condition is satisfied, this method creates a new forbidden tuple. In Figure \ref{fig:tuples_input_file}, the rule condition is fulfilled for parameter \textit{shape}, and it is evident that the new forbidden tuple is \textit{color=black}.

The \textbf{simplify} method finds the tuples that can be removed from the set while ensuring that the results of constraints validity will not be affected. Here, a forbidden tuple that contains another tuple from the set can be removed because any test satisfying a tuple must cover the subset of that tuple. In our example, the \textbf{derive} method finds tuple \textit{color=black}, which is a subset of each of the first three tuples. This means that the first three tuples can be removed.

A single call to the \textbf{derive} and \textit{simplify} methods cannot ensure that the transformation of constraints into forbidden tuples is complete because the \textbf{derive} method generates new implicit tuples that can be used for generating other implicit tuples. The varianter runs these methods in a loop, and the loop ends when the methods cannot change the forbidden tuples any more, thereby indicating that the tuple transformation is complete. Subsequently, the varianter generates a complete forbidden tuple set and cleans the combination matrix using the forbidden combinations. Figure \ref{fig:tuples_after_preprocessing} presents the set of tuples from Figure \ref{fig:tuples_input_file} after this process, and the cleaned combination matrix is illustrated in Figure \ref{fig:cleaned_combination_matrix}.

\begin{figure}
    \begin{verbatim}
    {{color=black, shape=square},
    {color=black, shape=triangle},
    {color=black, shape=circle},
    {color=gold, coating=cathodic},
    {material=aluminum, color=gold}}
    \end{verbatim}
    \caption{Forbidden tuples form the input file}
    \label{fig:tuples_input_file}
\end{figure}

\begin{figure}
    \begin{verbatim}
    {{color=black},
    {color=gold, coating=cathodic},
    {material=aluminum, color=gold}}
    \end{verbatim}
    \caption{Forbidden tuples after preprocessing}
    \label{fig:tuples_after_preprocessing}
\end{figure}

\begin{figure}
\scriptsize
    \bordermatrix{\text{ }&(0,0)&(0,1)&(0,2)&(1,0)&(1,1)&(1,2)&(2,0)&(2,1)&(2,2)\cr
                    (0,1)&-1 &-1 &-1 & 0 & 0 & 0 & 0 & 0 & 0\cr
                    (0,2)&-1 &-1 &-1 & 0 & 0 & 0 & 0 & 0 & 0\cr
                    (0,3)&-1 &-1 &-1 & 0 & 0 &-1 & 0 & 0 & 0\cr
                    (0,4)&-1 &-1 &-1 & 0 &-1 &-1 & 0 & 0 &-1\cr
                    (1,2)& 0 & 0 & 0 & 0 & 0 & 0 & 0 & 0 & 0\cr
                    (1,3)& 0 & 0 & 0 & 0 & 0 & 0 & 0 & 0 & 0\cr
                    (1,4)& 0 & 0 &-1 & 0 & 0 &-1 & 0 & 0 &-1\cr
                    (2,3)& 0 & 0 & 0 & 0 & 0 & 0 & 0 & 0 & 0\cr
                    (2,4)& 0 & 0 &-1 & 0 & 0 &-1 & 0 & 0 &-1\cr
                    (3,4)& 0 & 0 &-1 & 0 & 0 &-1 & 0 & 0 &-1\cr}
    \caption{Cleaned combination matrix}
    \label{fig:cleaned_combination_matrix}
\end{figure}

%%%%%%%%%%%%%%% Bestoun 12 December

\subsection{Generation of an initial solution}

Post the cleanup of the combination matrix based on the generated forbidden tuples, the combination matrix comprises valid combinations that must be covered. The varianter generates an initial solution that is the set of test cases that cover the valid combinations (i.e., t-tuples). At this stage, the aim is to find any solution that covers all combinations, without considering its optimality. Generation of the initial solution involves processing the test cases iteratively until the combination matrix is fully covered. The pseudocode to generate the initial solution is presented in Algorithm \ref{fig:initialization_pseudocode}.

\begin{algorithm}

 \KwIn{combinationMatrix}
 \KwOut{Testcases}
 
     solution = [ ]
     
     combinationMatrix = computeCombinations(data, tValue)
     
     \While{uncoveredNumber != 0}{
     
     \If{$uncoveredNumber > coveredMoreThanOnesNumber$}{
     
        $testCase_1 = randomTestCase$
        
        $testCase_2 = randomTestCase$
        
        \If{$distance(testCase_1,solution)>distance(testCase_2,solution))$}{
        
           $testCase = testCase_1$
            
        }\Else{
        
            $testCase = testCase_2$
            
        }
        
     }\Else{
     
        $testCase = coverMatrix()$
        
     }
     
     combinationMatrix.coverCombinations(testCase)
     
     solution.add(testCase)
     
     }
     
     Return $solution$

    \caption{Solution initialization}
    \label{fig:initialization_pseudocode}
\end{algorithm}

\begin{figure}
\scriptsize
\centering
    \includegraphics[width= 1.4 in]{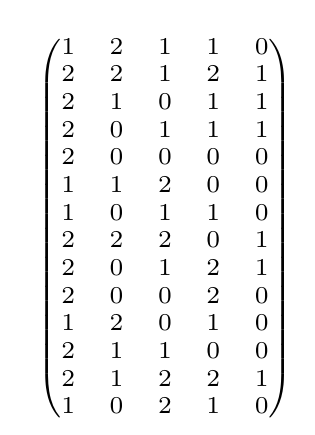}
    \caption{Initial solution}
    \label{fig:Initial solution}
\end{figure}

The random search mechanism of the Avocado plugin is, in fact, not entirely random. Here, we followed the "Hamming distance" mechanism presented in \cite{GONZALEZHERNANDEZ201517}. Only the first test cases are generated randomly, whereas the remaining are created from the first ones. During the random search, the varianter first creates two random test cases that are valid as per the constraints. Subsequently, for each of these test cases, the varianter computes the Hamming distance from an already found solution and chooses the one with largest Hamming distance. Then, this test case is added to the solution. The Hamming distance between the test case and the initial solution is the sum of the Hamming distances between the test case and each test case of the initial solution. The random search generates new test cases until the number of the uncovered tuples is smaller than the number of tuples covered by more than one test case. At this point, the search process becomes inefficient because the probability of covering a new tuple is smaller than the probability of covering an already covered tuple. In this scenario, the varianter must complete and fill the initial solution by a discrete method for covering the tuples. This method uses all uncovered tuples from the combination matrix, chooses the largest disjunct set from them, which will be covered by the next test case. Some of the values from the new test case may be not defined by the disjunct set. These values are filled randomly by the varianter. This discrete method adds new test cases to the initial solution until all tuples are covered and the initial solution is complete. Figure \ref{fig:Initial solution} shows a sample solution where the rows denote test cases and the columns denote parameters of the SUT.

\subsection{Improvement of the solution}

Eventually, the set of test cases that covers all t-tuples of parameter values created by the varianter becomes available. This set can be used for testing the SUT. However, this set of test cases is not optimal and it is possible to cover the combinations with a smaller set of test cases. The aim of the CIT varianter is to find the smallest possible valid solution at a specified time. However, Avocado does not aim to create new bounds of CAs.

To this end, the next stage of the varianter algorithm is to optimize the size of the initial set of test cases while maintaining the coverage of all tuples. Here, the CIT varianter uses the mixed neighborhood tabu search (MNTS) algorithm that is newly developed by us \cite{Hasan2019GenerationAA}. This algorithm is based on the randomized tuning of the solution that leads to a better solution. As shown in Figure \ref{fig:Initial solution}, MNTS represents the solution as a matrix of size $N \times M$, where each row represents one test case and each column represents one input parameter of the SUT. MNTS aims to delete the specified number of rows from the solution matrix. Then, using the randomized algorithms, the new solution is modified to cover the combination matrix. In other words, the level of tuple coverage is maintained, covering all tuples with a smaller number of rows (i.e., test cases) in the matrix. Algorithm \ref{fig:improvement_pseudocode} provides the pseudocode and the procedure of this algorithm.

\begin{algorithm}

 \KwIn{solution, combinationMatrix}
 \KwOut{Testcases}
     M = 600\\
     newSolution = solution.copy()\\
     combinationMatrix = computeCombinations(data, tVvalue)\\
     deletedRow = newSolution.deleteRow()\\
     combinationMatrix.uncoverCombinations(deletedRow)\\
     \For{$i=0$ to M, $i+1$}{
     algorithm = choose(1,2,3)\\
     newSolution = algorithm(newSolution)\\
     combinationMatrix.updateCoverage(newSolution)\\
     \If{uncovered == 0}{
        Return newSolution\\
     }
     }
     Return solution\\
    
    \caption{Solution improvement}
    \label{fig:improvement_pseudocode}
\end{algorithm}

\begin{comment}
\begin{figure}
\begin{lstlisting}
function compute_better_solution(solution,combination_matrix){
    M = 600
    new_solution = solution.copy()
    delete_row = new_solution.delete_row()
    combination_matrix.uncover_combinations(delete_row)
    final_matrix = []
    while uncovered != 0{
        N = choose(1,2,3)
        new_solution = N(new_solution)
        combination_matrix.update_coverage(new_solution)
        if M == 0{
            return solution
        }
    }
    return new_solution
}
\end{lstlisting}
    \caption{Solution improvement pseudocode based on MNTS}
    \label{fig:improvement_pseudocode}
\end{figure}
\end{comment}

The algorithm initially deletes a random row from the solution matrix and uncovers the t-tuples of the deleted test cases. Uncovering the t-tuples means decreasing by 1, every cell in the combination matrix that is covered by the deleted test case. The solution that is not complete at this point must be modified to cover the entire combination matrix. For every modification to the solution matrix, one of the three modification functions N\textsubscript{1}, N\textsubscript{2}, and N\textsubscript{3} is randomly chosen with probability 0.1, 0.1, and 0.8, respectively. This probability ratio is based on the experiments described in \cite{GONZALEZHERNANDEZ201517}. Each modification function modifies the solution matrix to create a new matrix, which is achieved by changing the numbers in the randomly selected cells. The modification with the best coverage of the combination matrix is selected as a new solution matrix for other modifications. N\textsubscript{1} randomly selects a cell of the solution matrix and makes all possible changes to the number in the cell. N\textsubscript{2} randomly selects a column in the solution matrix and changes each cell in this column. N\textsubscript{3} randomly selects one of the uncovered combinations to form the combination matrix and changes every row in the solution matrix to cover this combination. When the modified solution covers the entire combination matrix, the algorithm executes the initial step of deleting another row from the matrix and continues. The algorithm defines the constant, \textit{m}, that represents the maximum number of modifications to the solution. We have chosen the maximum number of modifications based on our experimental experience and preferences. However, the user has the option to change this number based on the application preferences. When the number of modifications reaches \textit{m}, it is concluded that a better solution cannot be found, and the last complete solution is treated as the best solution.

\section{Conclusion}\label{conclusion}

In this paper, we presented a CIT plugin that promotes a fully automated and comprehensive strategy to use the capabilities of CIT on an industrial scale. The framework is implemented with a well-known and maintained automated testing framework called Avocado. Avocado, which is an open-source tool has recently seen application in several use cases. This paper reports our recent research efforts to develop a successful flexible constraint handling strategy to resolve the constraints in the final test suites. We combined and used the best practices in literature, including our past research efforts to implement this tool. The tool is maintained by the Avocado development team, and several directions for future implementations have been identified by the community because it is an open-source project. A significant direction for improvement is how to deal with large input parameters, values, and constraints on a mega-scale.

\section{Acknowledgement}

This research is funded by Red Hat Czech s.r.o. as a collaboration project within the Avocado testing framework and the support of the OP VVV funded project CZ.02.1.01/0.0/0.0/16\_019/0000765 "Research Center for Informatics"

\bibliographystyle{IEEEtran}
\bibliography{sample}

\end{document}